\documentstyle[aps,manuscript,epsf]{revtex}  % DON'T CHANGE %

\newcommand{\as}{$\alpha_S$ }
\newcommand{\amsbar}{\alpha_{\overline{\rm MS}}}

\begin{document}

% --- TITLE

\title{\vskip -3cm \makebox[11cm]{}{\normalsize WUB 99-13} \\
\vspace{-0.5cm}\makebox[11cm]{}{\normalsize HLRZ 99-20} \\ 
{$\alpha_S$ from $\Upsilon$ Spectroscopy with dynamical Wilson Fermions}}

\author{SESAM-Collaboration} 

\author{A.~Spitz$^{\rm a,b}$\thanks{email:spitz@theorie.physik.uni-wuppertal.de}, H.~Hoeber$^{\rm a}$, N.~Eicker$^{\rm
    a,b}$, S.~G\"usken$^{\rm a}$, Th.~Lippert$^{\rm a}$, \\
  K.~Schilling$^{\rm a,b}$, T.~Struckmann$^{\rm a}$,
    P.~Ueberholz$^{\rm a}$, J.~Viehoff$^{\rm b}$}

\address{$^{\rm a}$Fachbereich Physik, Bergische Universit\"at Wuppertal,\\
         D-42097 Wuppertal, Germany}
\address{$^{\rm b}$John von Neumann-Institut f\"ur Computing,\\
         Forschungszentrum J\"ulich, D-52425 J\"ulich, Germany}

\date{11.06.1999}
\maketitle

% --- ABSTRACT

\begin{abstract}
  We estimate the QCD coupling constant from a lattice calculation of the
  bottomonium spectrum. The second order perturbative expansion of the
  plaquette expectation value is employed to determine $\alpha_S$ at a scale
  set by the 2S-1S and 1P-1S level splittings. The latter are computed in
  NRQCD in a dynamical gauge field background with two degenerate flavours of
  Wilson quarks at intermediate masses and extrapolated to the chiral limit.
  Combining the $N_f=2$ result with the quenched result at equal lattice
  spacing we extrapolate to the physical number of light flavours to find a
  value of $\amsbar^{(5)} (m_Z) = 0.1118(17)$. The error quoted
  covers both statistical and systematic uncertainties in the
  scale determination. An additional $5\%$ uncertainty comes from the
  choice of the underlying sea quark formulation and from truncation
  errors in perturbative expansions.
\end{abstract}

% --- INTRODUCTION

\section{INTRODUCTION}
A precise knowledge of the strong coupling constant is of central
importance both for strong-interaction phenomenology and for stringent
tests of the standard model as a whole.  
There has been significant
progress in determining $\alpha_S$ and its running during the last
years: Measurements are available for a large number of different
reactions covering energies up to 190 GeV. Although `the' global
average has remained nearly unchanged the uncertainties of single
measurements as well as the scatter between them has been strongly
reduced. The clustering of results into groups of high-energy and
low-energy data which was observed back in 1995 has completely
disappeared and the various results now nicely align around an average
of $\alpha_{\overline{\rm MS}}(m_Z)\sim 0.119$ \cite{bethke98}.
While the majority of determinations of the strong coupling relies on
perturbative expansions of high-energy cross sections, lattice
simulations of QCD provide, in principle, a nonperturbative access to 
determine \as from low-energy quantities (a review of lattice
methodology is presented in \cite{weisz96}). In fact, a viable scheme
along these 
lines has been demonstrated with the Schr\"odinger functional technique in the
context of the quenched approximation, $N_f = 0$~\cite{luscher94}.  An
alternative approach is to consider perturbative expansions of suitable
short-distance quantities and to determine their vacuum expectation values
from simulations. In this manner Monte Carlo data can be utilized to extract
estimates for \as at a scale which is provided by the lattice spacing,
$a$ \cite{elkhadra92}.
This procedure lends itself easily to the setting of full QCD, given that
sufficient samples of QCD vacuum configurations are available as demonstrated
some time ago in ref. \cite{davies95}. In the meantime, the statistical
precision of the resulting lattice estimates of \as in $N_f = 0$ and $2$
theories has been increased such as to allow for an extrapolation in the
number of dynamical flavours to $N_f=3$.

Although, a priori, any mass or energy splitting calculated in a lattice
simulation can be used to set the lattice spacing, $a$, in practice one
clearly prefers those choices which are least prone to systematic errors.
Spin-averaged radial and orbital angular momentum splittings in heavy
quarkonium bound states, in particular in bottomonium, are favourite
candidates \cite{lepage92a}. They are largely insensitive to the heavy quark
mass: experimentally one finds the spin-averaged mass splittings between 1P
and 1S levels as well as 2S and 1S levels to be nearly equal for the
$\Upsilon$ and $\Psi$ bound states.  Moreover, one may infer
phenomenologically, that they depend only mildly on the light quark masses:
assuming characteristic gluon momenta exchanged between quarks inside
the $\Upsilon$ to be $O(1 {\rm GeV})$, heavy flavours will be negligible in
virtual quark loops whereas the masses of light quarks are equally small
compared to this momentum scale so that their exact values as well as the
differences between non-strange and strange quark masses are likely to be of
little importance.  These arguments imply that {\it (a)} there is no need to
tune the bare heavy quark mass precisely and {\it (b)} to a good
approximation the contribution of light quarks can be mimicked by considering
three light flavours of average mass \cite{shigemitsu96}. The latter
feature is a major benefit in lattice simulations which are technically
restricted to rather large quark masses and generally rely on chiral
extrapolations.

The {\it bonus} in choosing spin-averaged level splittings in bottomonium to
set the scale stands against the {\it malus} that the direct simulation of
heavy quark dynamics employing standard relativistic actions suffers from
serious discretization errors. Among several alternatives to circumvent this
problem, the nonrelativistic effective theory for QCD (NRQCD) \cite{lepage92}
provides the most appropriate technique for simulating systems containing a
b-quark.  The NRQCD Lagrangian is written as a series of operators ordered
according to powers of the mean-squared heavy-quark velocity. Each operator
introduces a new coupling constant which is determined by perturbatively
matching the theory to QCD. The decoupling of quark and antiquark fields in
NRQCD is equivalent to integrating out the heavy quark mass and, as a
consequence, the amount of lattice spacing errors is governed by the
characteristic momentum exchanged between heavy quarks, $O(a\Lambda_{\rm
  QCD})$, rather than  by the underlying quark mass scale.

The realistic inclusion of vacuum polarization effects constitutes a serious
hurdle in today's lattice simulations of QCD, in particular for Wilson
fermions. One should remember, that the
state-of-the-art updating algorithms require an even number of
degenerate sea quarks. As a consequence the strong coupling cannot be
determined on field configurations with three kinds of active sea quarks, $u$,
$d$, $s$; rather one has to resort to an extrapolation in the number of active
flavours. Using the staggered discretisation for the dynamical fermions,
Davies and coworkers\cite{davies95} presented a careful
analysis of \as (including the various sources of error). Consistent
results, although with substantially larger errors, have been quoted
by the authors of \cite{wingate95,aoki95} who also use 
staggered sea quarks but heavy Wilson valence quarks instead of
nonrelativistic quarks.

In this paper, we shall apply the methods of \cite{davies95} to the
case of dynamical Wilson fermions.  They carry different 
finite-a errors (compared to staggered fermions) that affect both the
nonperturbative results for the bottomonium splittings and the plaquette
expectation value as well as the perturbative expansion of the plaquette.  We
shall improve on previous error analyses by studying the quark mass dependence
of quarkonium splittings; hence we shall be able to reduce the uncertainty
due to the light quark mass scale.

The paper is organized as follows. In Sec. \ref{sec:simulation} we
briefly recall the set-up of SESAM's hybrid Monte Carlo simulation of
lattice QCD with dynamical Wilson fermions as well as the relevant
issues of the computation of nonrelativistic propagators from a
lattice NRQCD action. We proceed with a presentation of spectroscopy
results in Sec. \ref{sec:spectroscopy}. Since within the $\Upsilon$
system the singlet 
states $\eta_b$ and $h_b$ have not been observed in experiment yet, we
have to deviate from the suggestion to use fully spin-averaged splittings
and have to choose $2^3S_1-1^3S_1$ and $^3\bar P-1^3S_1$ to determine the
lattice spacing. Here, $^3\bar P$ denotes the spin average of
$^3P_J$. Therefore the discussion of fit results and 
systematic errors in Sec. \ref{sec:spectroscopy} covers radial
splittings as well as 
spin splittings. In Sec. \ref{sec:alphaP} we determine the quenched
and unquenched plaquette couplings and perform the extrapolation in
$N_f$. We conclude in Sec. \ref{sec:discussion} with a discussion of
the resulting value of $\alpha_S$ in the continuum $\overline{\rm MS}$ scheme.

% --- SIMULATION SET-UP

\section{SIMULATION SET-UP}
\label{sec:simulation}
The present work is based on ref. \cite{spitz98} and can be described
as the final analysis of SESAM lattices with 
increased statistics, both with respect to the number of configurations and
NRQCD propagator inversions, to reduce the error.

\subsection{Gauge Fields}
We have performed a hybrid Monte Carlo simulation of full QCD at $\beta=5.6$
with two degenerate flavours of dynamical standard Wilson fermions. A single
lattice size of $L^3\times T = 16^3\times 32$ is used. It corresponds to a
physical box of 1.2-1.4 fm in the spatial direction (depending on the
experimental quantity used to set the scale) which is sufficiently large to
exclude finite volume effects on the bottomonium ground state and the first
radial and orbital angular momentum excitation. We generated configurations at
three different values of the sea quark hopping parameter, $\kappa$, each
sample consisting of 5000 trajectories from which 200 decorrelated vacuum
configurations are chosen. Table \ref{tab:simsetup} lists the simulation
parameters. For details of the HMC run and subtleties concerning
autocorrelation studies we refer the reader to \cite{light98}. The quenched
run uses an overrelaxed Cabbibo-Marinari heat bath update, thermalised with
2000 sweeps, and measurements are performed on configurations separated by 250
sweeps.

\subsection{Some NRQCD Prerequisites}
 Compared to our intermediate  results presented 
in \cite{nrqcd97},  we have significantly increased our
statistics at $\kappa = 0.1575$ and $\kappa=0.1570$ corresponding to the two
lighter quark masses.  Heavy quark propagators are calculated in the
nonrelativistic approximation.  We have implemented the NRQCD action at
`next-to-leading order' with spin-independent operators of $O(m_b v^2)$,
$O(m_b v^4)$ and spin-dependent terms of $O(m_b v^4)$ and $O(m_b v^6)$
included. The radiative corrections to the coefficients of these interaction
terms are not exactly accounted for. Rather we use the mean field prescription
to cover their main effect and choose  the tadpole parameter computed from
the mean link in Landau gauge. Unlike for light hadrons, in lattice
simulations of heavy quarkonium bound states one can efficiently increase
statistics by computing propagators from different, widely separated source
points on the same configuration.  Hence we have averaged the data from up to
12 point sources for the $N_f=2$ lattices. Note, that we do not use a
multi-source, but evolve each starting point separately. In the quenched runs
we restrict ourselves to 4 source points as new configurations can be
generated with comparatively little effort.  To project on radially excited
and orbital angular momentum states some care is needed in
choosing an approriate smearing procedure. The method of choice for
bottomonium is a fixed-gauge wavefunction smearing technique. For the
$\Upsilon$ and $\eta_b$ we calculate a $4\times 4$ matrix of correlators with
four different smearings at source and sink, $sc/sk = l,1,2,3 $, corresponding
to a point source (l), the ground state (1), the first (2) and second (3)
excited states, respectively.  For the $L=1$ states we restrict ourselves to
the ground state and the first excitation as signals deteriorate. The smearing
functions that we use are the wavefunctions calculated in Ref. \cite{bali97}.
With all the coefficients in the action set to their tree level values the
only parameter left besides the gauge coupling is the bare heavy quark mass,
$am_b$.  We did not tune the bare b-quark mass but kept $am_b = 1.7$
throughout the simulation. This value reproduces the correct $\Upsilon$ mass
in the quenched approximation \cite{nrqcd97} and it turns out to be adequate
in the full theory, too, leading to kinetic masses $m_{\rm kin}(\Upsilon) =
9.97(28),9.63(24),9.68(27)\; {\rm GeV}$ for $\kappa = 0.1560,0.1570,0.1575$,
respectively.

% --- BOTTOMONIUM SPECTROSCOPY

\section{BOTTOMONIUM SPECTROSCOPY}
\label{sec:spectroscopy}

\subsection{Fit Results}
\begin{figure}
\begin{center}
  \begin{minipage}{8cm}
    \epsfxsize=8cm
    \epsfbox{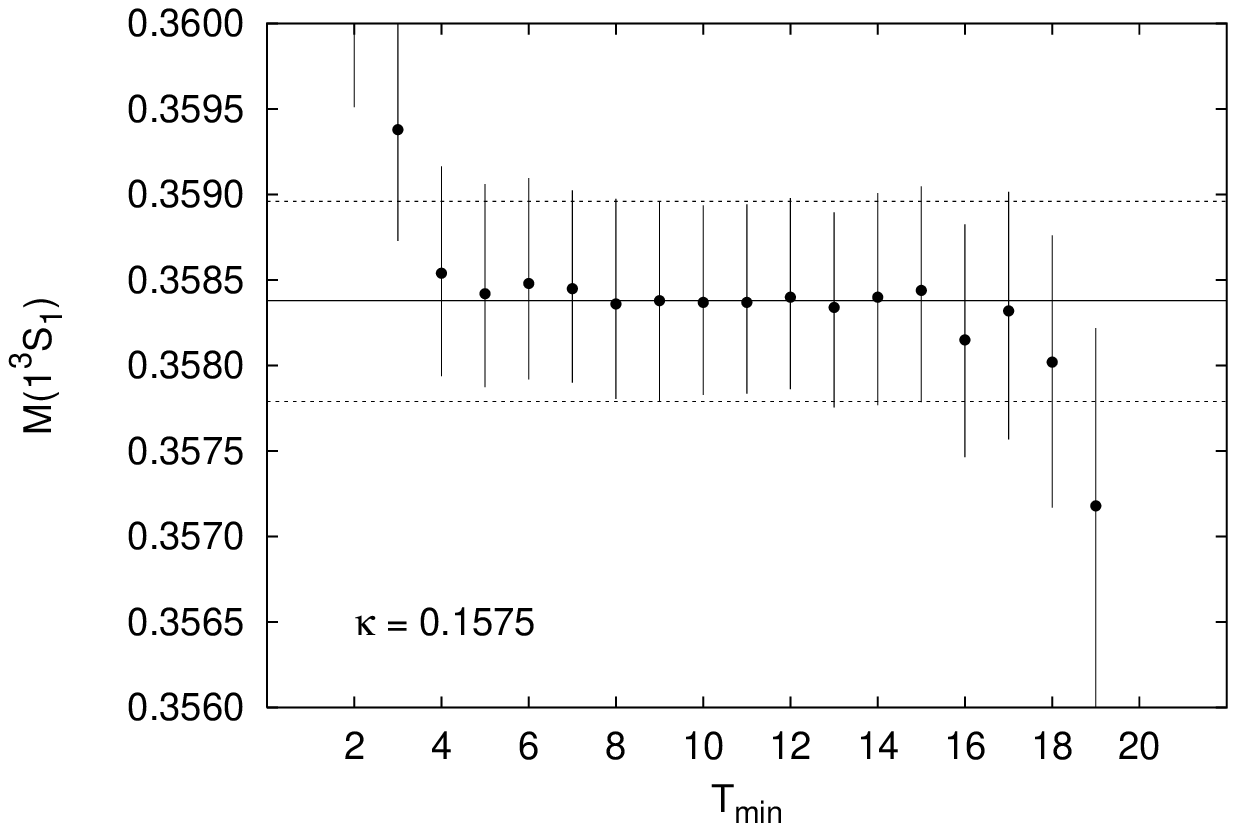}
  \end{minipage}
  \begin{minipage}{8cm}
    \epsfxsize=8cm
    \epsfbox{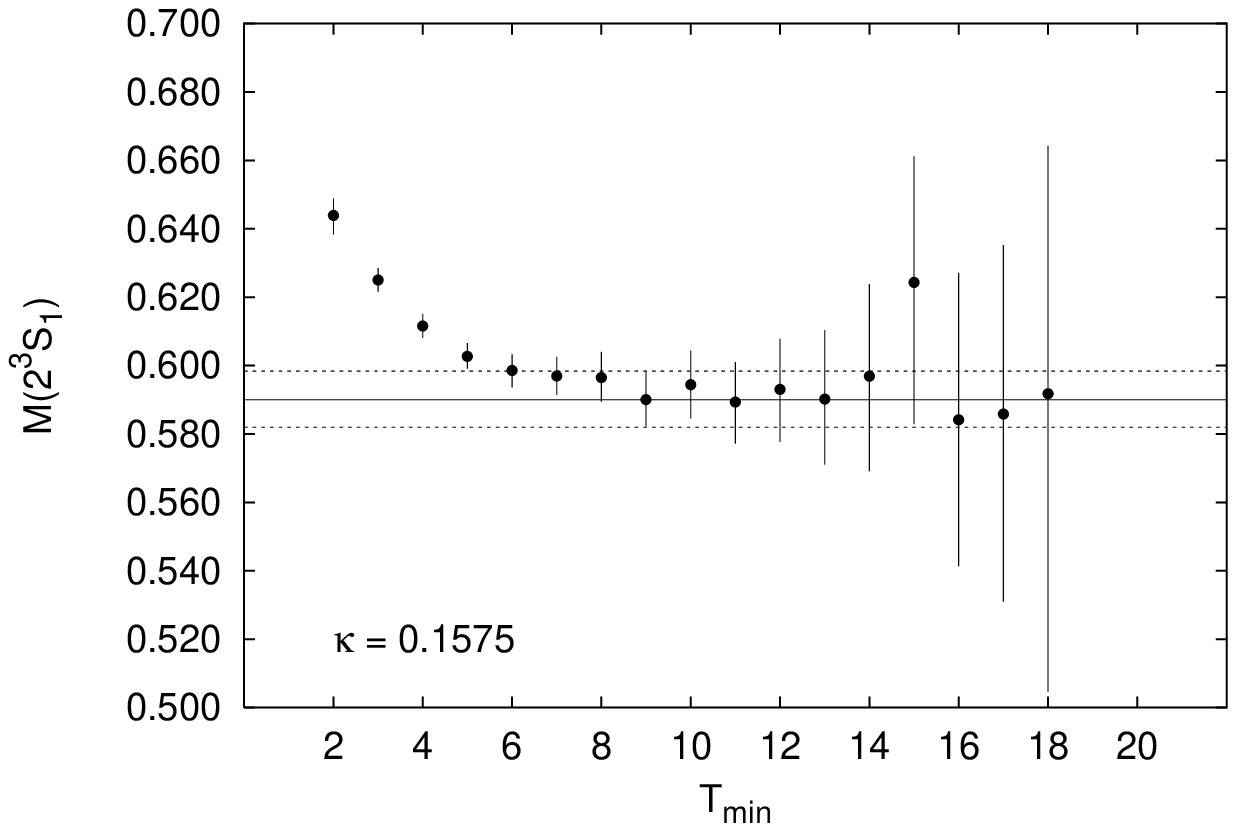}
  \end{minipage}
  \\
  \begin{minipage}{8cm}
    \epsfxsize=8cm
    \epsfbox{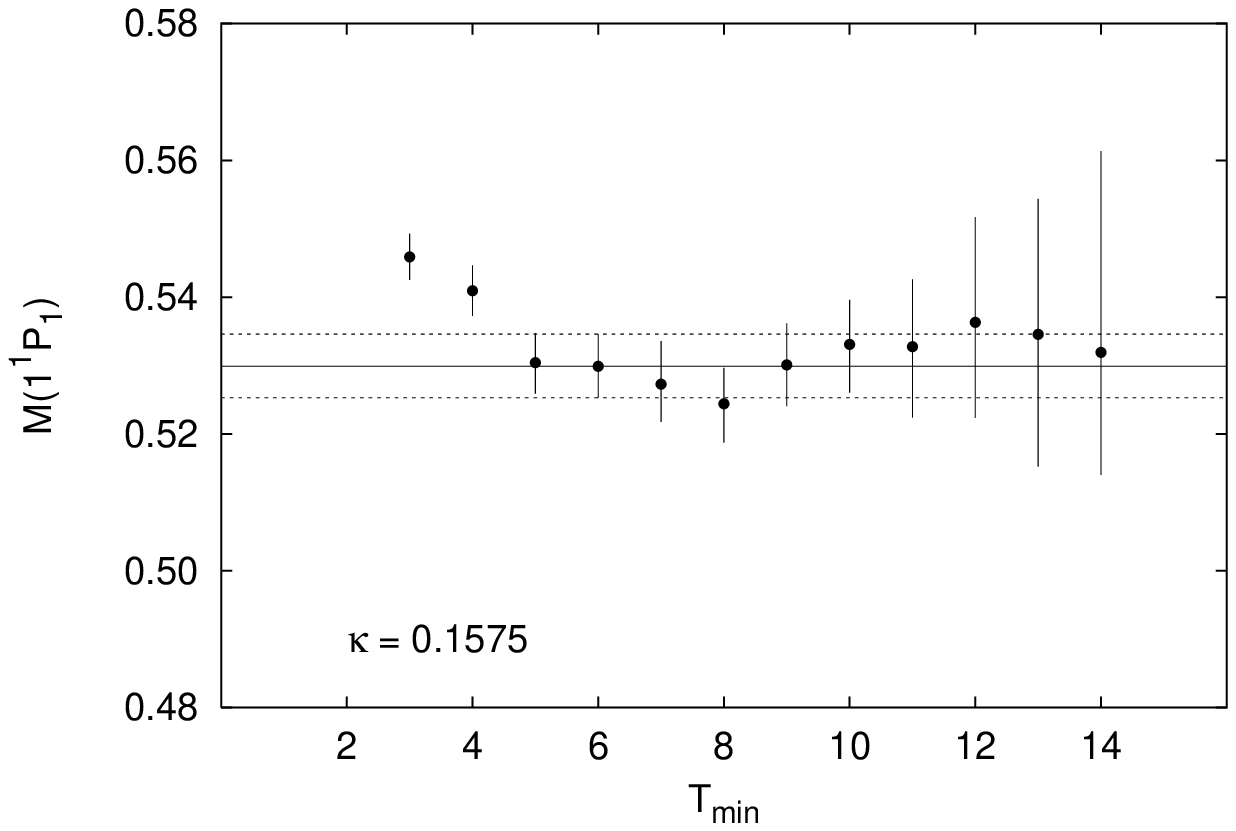}
  \end{minipage}
  \begin{minipage}{8cm}
    \epsfxsize=8cm
    \epsfbox{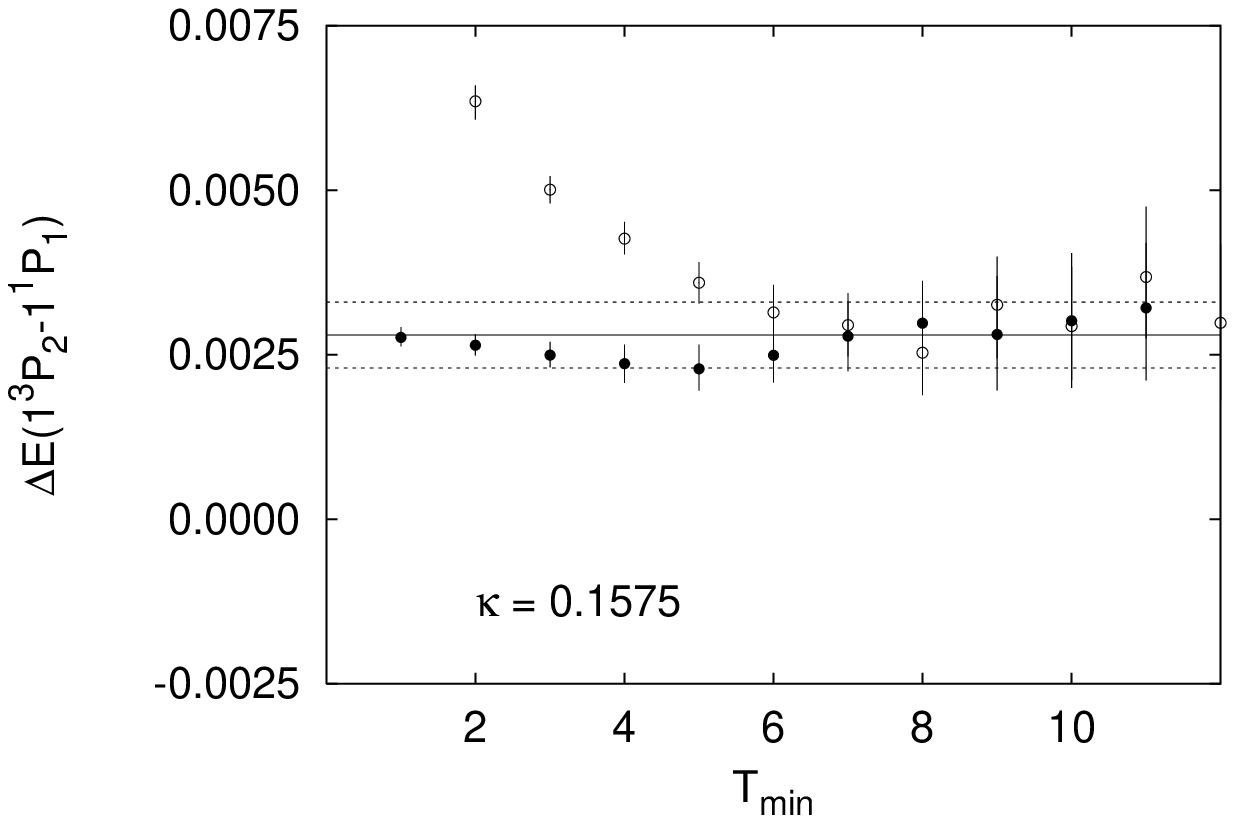}
  \end{minipage}
\end{center}
\caption{\sl Global masses as a function of $T_{\rm min}$ with $T_{\rm
    max}=30$. Lines mark the selected fit value (solid) and its error
    bands (dotted). For the spin-splitting (lower right plot)
    smeared-local (open symbols) and smeared-smeared data (filled
    symbols) are plotted. \label{fig:tminplot}}
\end{figure}

We determine the triplet-S state $1^3S_1$
and its first radial excitation as well as the singlet-P ground state
$1^1P_1$ by a simultaneous fit of two source-smeared
correlators to a double-exponential ansatz
\begin{equation}
  C_{\rm sc,l}(T) = b_{\rm sc,l}^1 {\rm e}^{-a E_1 T} + b_{\rm
    sc,l}^2 {\rm e}^{-a E_2 T}\; \quad sc = 1,2 \; .
\end{equation}
These fits yield the cleanest signals and are very stable. We varied the fit
interval over a considerable range as illustrated in Fig. \ref{fig:tminplot}
for the sample with the lightest sea quark content. The analysis reveals clear
plateaus in global masses for each meson correlator. Since fits to
smeared-local correlators usually tend to overestimate the plateau value, we
have varied the fitting procedure to check for the stability of the results.
First, we have used correlators which are smeared both at source and sink in a
two-exponential fit. Sink smearing produces noisier signals. Nevertheless,
ground state energies are obtained with similar accuracy and yield consistent
results. Excited states, however, exhibit
significantly larger errors compared to the smeared-local data.  Second, we
have applied fits involving two smeared-local correlators but three
exponentials. The third exponential is meant to account for contaminations
from higher radial excitations and to provide a more reliable estimate of the
2S state. The results are in very good agreement with those obtained from two
exponentials, thus providing  confidence in the radially excited level.
Since both singlet resonances, $\eta_b$ and $h_b$, have not been observed
experimentally we have to recourse to the known splittings $2^3S_1 - 1^3S_1$
and $\bar P - 1^3S_1$ to set the lattice scale. Here, $\bar P$ denotes the
spin-average triplet-P state: $^3\bar P = \frac{1}{9}\left(^3P_0 + 3\cdot^3P_1
  + 5\cdot^3P_2\right)$. To obtain $^3\bar P$ we have to determine the P fine
structure which is accomplished by single exponential fits to the ratio of two
correlators
\begin{equation}
  C_1(T)/C_2(T) = A\;{\rm exp}(-\Delta E_{12} T)\; , \quad \Delta E_{12} =
  E_1 - E_2 \; .
\end{equation}
This way we compute the splittings of $^3P_J$ relative to $^1P_1$ for
$J=0,1,2$ which are then combined to form the spin average.
In the case of smeared-local correlators the
plateau sets in only at rather large times. For P-wave states these
can hardly be reached before the signal is drowned into noise. Results
quoted for spin splittings have therefore been taken from
smeared-smeared correlators. These run into plateaus at early times
where they differ considerably from their smeared-local
counterparts as illustrated  in Fig. \ref{fig:tminplot}. Table
\ref{tab:final_ro} summarizes the relevant fit results. Errors are
taken from 300 bootstraps.

\subsection{Extrapolation in the light quark mass}
\begin{figure}
\begin{center}
  \begin{minipage}{8cm}
    \epsfxsize=8cm
    \epsfbox{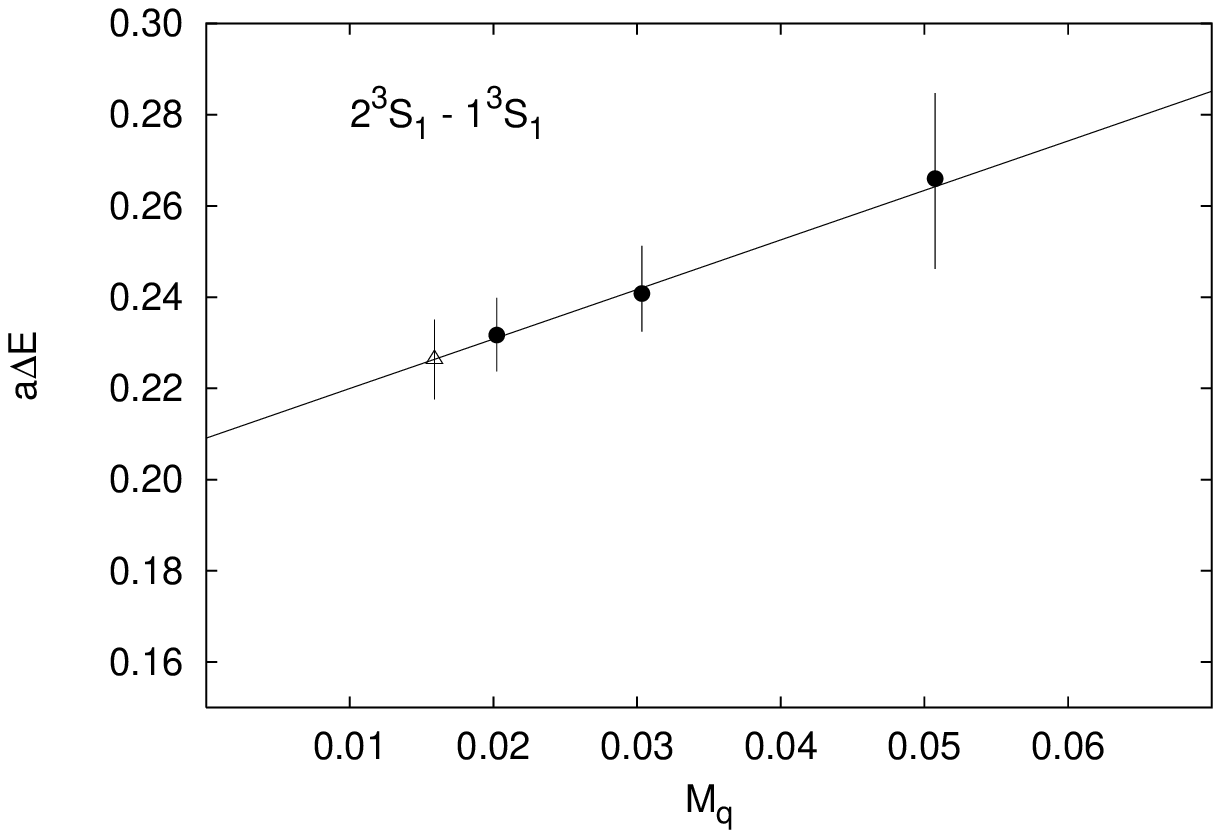}
  \end{minipage}
  \begin{minipage}{8cm}
    \epsfxsize=8cm
    \epsfbox{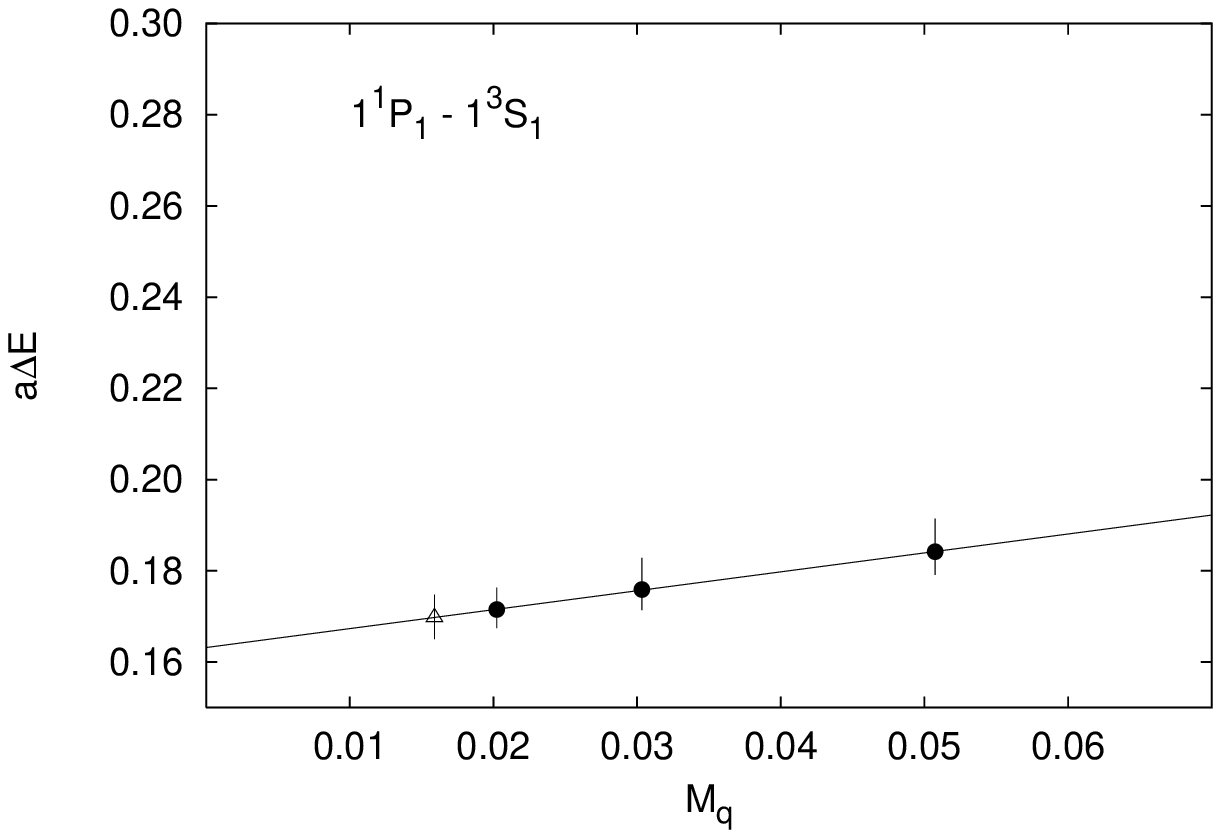}
  \end{minipage}
\caption{\sl Extrapolation of the splittings between the $\Upsilon$ ground
  state and its radial and orbital angular momentum excitation in the
  dynamical quark mass. The open triangle denotes the value at $\tilde
  m$.\label{fig:extrap}}
\end{center}
\end{figure}

The hopping parameters that we have chosen correspond to quark masses
in the range $m_s/2 \le m_q \le m_s$ where $m_s$ denotes the strange
quark mass. 
According to \cite{grinstein96} one expects the energy splittings to depend
linearly on the light quark mass, $m_q$, in lowest order of a chiral
expansion. Hence we choose the simple ansatz
\begin{equation}
    a\Delta E = a\Delta E_0 + c\, am_q 
\end{equation}   
to extrapolate in $m_q$. As illustrated in Fig. \ref{fig:extrap} the 
splittings considered here exhibit a very mild dependence on the sea quark
mass. In fact, one cannot distinguish their values at $a\tilde m \equiv
am_s/3 \sim 0.0159$ and our smallest simulated mass.\footnote{Note that $\kappa_s$ determined using the $K$ mass disagrees
  with the value calculated from $K^*$ or $\Phi$. We use an average
  here since the exact value of $\kappa_s$ is clearly not important
  in the present analysis.}
As outlined in the introduction $\tilde m$ is a reasonable average
quark mass provided the strange-nonstrange mass difference can be
neglected for the lowest $\Upsilon$ bound states.
If extrapolated down to the physical light quark mass the splittings
increase modestly. The parameters of uncorrelated linear fits are
listed in Table \ref{tab:extrap_results} together with the splittings at
$\tilde m$ and $m_l$. We conclude that within the accuracy of our data
no deviation from linear behaviour is found and we emphasize that sea
quark mass dependences are smaller by one order of magnitude compared
to the light hadron sector \cite{light98}.

We may choose the difference in mean values of $a\Delta E(m_s/3)$ and
$a\Delta E(m_l)$ as an upper limit of the uncertainty in the sea quark
mass: $1^3\bar P - 1^3S_1$ exhibits a 4\% decrease, whereas $2^3S_1 -
1^3S_1$ is affected by 7\%. From the experimental bottomonium
splittings $\Upsilon ' - \Upsilon$ = 0.5629 GeV and $\bar\chi -
\Upsilon$ = 0.4398 GeV one determines the lattice scales summarized in
Table \ref{tab:spacings}. We use the average value of $a^{-1}$ at
$\tilde m$ to convert our lattice results into physical units. As
intended, it matches the (mean) value of the quenched lattice spacing
at $\beta=6.0$, so that unquenching effects can be studied.

\begin{figure}
\begin{center}
  \epsfxsize=16cm  
  \epsfbox{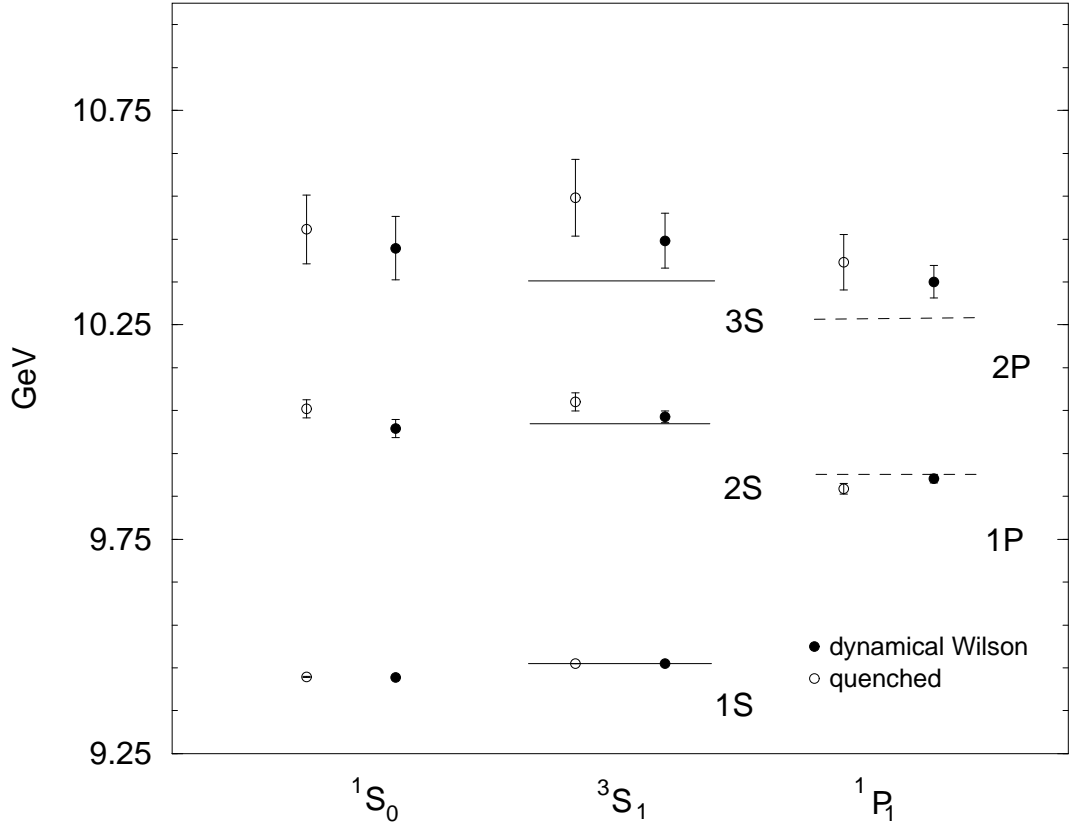}
\caption[Bottomonium spectrum - radial and orbital angular momentum
splittings]{\sl Bottomonium spectrum - radial and orbital angular momentum
  splittings. The $^3S_1$ ground state is constrained to match the
  experimental $\Upsilon$ energy. Open symbols denote quenched results, filled
  symbols $N_f=2$ results at $m_q = m_s/3$. Solid lines mark the experimental
  values, dashed lines the position of the spin-averaged $^3P_J$ states, which
  turn out to be nearly identical with the singlet-P
  estimates.\label{fig:bb_spectrum}}
\end{center}
\end{figure}

It is obvious from Fig. \ref{fig:bb_spectrum} that the quenched
spectrum does not reproduce the experimental spectrum  while the
$N_f=2$ data is in much closer agreement. This behaviour can be
ascribed to the different running of the coupling in both theories and
translates into a discrepancy between the plaquette coupling
as determined from the 2S-1S and 1P-1S splittings in the quenched
approximation as we shall see below.  

\subsection{Error Estimates}

Besides the dependence on the sea quark mass several error sources
deteriorate the accuracy of the energy splittings and hence affect the
precision in scale determination:

Inherent to the nonrelativistic approach one is faced with higher order
relativistic corrections due to the truncation of the effective NRQCD
Lagrangian of choice and to the incomplete renormalization of coefficients in
the velocity expansion. Having included spin-independent terms through
$O(m_bv^4)$ we expect that spin-averaged splittings will receive a 1\%
correction from higher orders (interactions of $O(m_bv^6)$). But this is just
the order of magnitude incurring by second-order spin-dependent effects! Hence
we may estimate the truncation error by switching off the latter.  By
inspection of Table \ref{tab:lonlo} we find  that this procedure leaves
 the energy levels unaltered.
A ratio fit to the S hyperfine splitting, however, reveals a 15\% decrease
when $O(m_bv^6)$ interactions are switched on as is expected from power
counting. The impact of the
incomplete renormalization of the NRQCD expansion coefficients is
exposed by changing the tadpole improvement factor from the Landau-link
prescription, $u_0^L\equiv\langle \frac{1}{3}{\rm Tr} U_{\mu}\rangle_{\rm LG}$,
to the plaquette prescription, $u_0^P \equiv \sqrt[4]{\langle\frac{1}{3}{\rm
    Tr} U_{\mu\nu}\rangle}$. It turns out to be of similar magnitude.
As a conservative estimate, we shall henceforth quote an additional
error of 0.10 in $a^{-1}$ to cover both effects and refer to it as the
uncertainty in the NRQCD expansion.

\begin{figure}
\begin{center}
  \epsfxsize=12cm
  \epsfbox{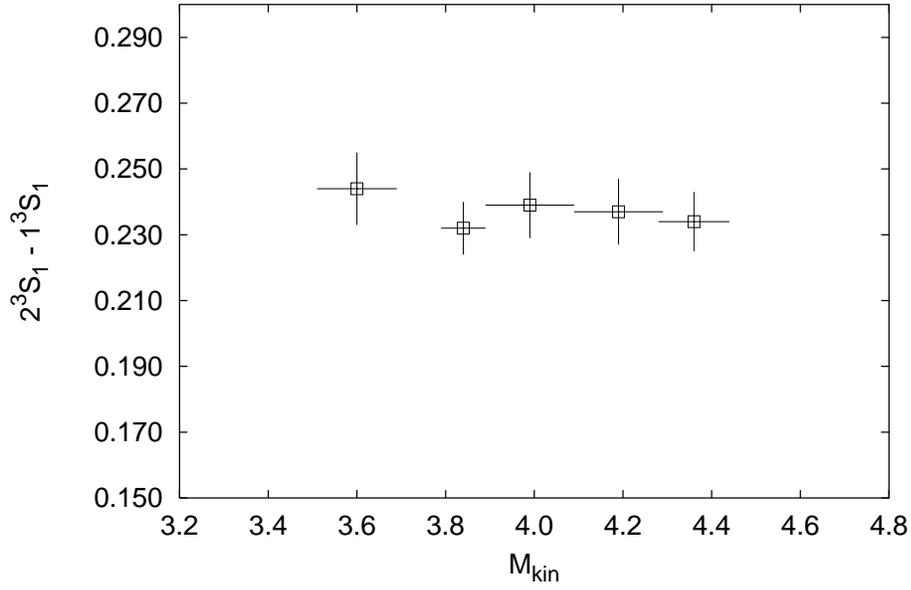}
\caption{Dependence of $2^3S_1-1^3S_1$ radial splitting on the kinetic
  mass.\label{fig:depkin}}
\end{center}
\end{figure}

We have explicitly calculated the dependence of the 2S-1S radial
splitting on the bare heavy quark mass. Of course, we expect very
little change  as we vary $am_b$. This is indeed confirmed as can be
seen from Fig. \ref{fig:depkin}, where we plot the triplet S radial
splitting as a function of kinetic mass. The latter is computed for
each bare mass value, $am_b = 1.6 - 2.0$, by
giving the meson a small amount of momentum and fitting the
nonrelativistic dispersion relation.

We have not checked for finite volume effects. We expect
them to be much smaller for quarkonia than for light mesons. Quenched
lattices of size $\sim$ 1.5 fm are found to be sufficient for the
lowest charmonium levels \cite{davies95}. Hence our results for 2S-1S and
1P-1S in bottomonium are safe. But, as pointed out in \cite{gunnar98},
higher radial excitations like 2P or 3S require a linear lattice extent
larger than 2 fm, even for bottomonium.

Finally, our results will be affected by various finite-a
errors. Note, that these cannot be removed by extrapolation, since
there exists no continuum limit to NRQCD. Instead the latter should be
viewed as being based on an effective action that is geared to obtain
physical results {\it at finite cut-off} only. Obviously, in oder to
make sure that the effective action approach is a useful tool one has
to ascertain that the spectrum is independent of the lattice spacing
within a certain window.  This then would establish that the matching
of the NRQCD action to QCD is sufficiently accurate.

For nonrelativistic b-quarks discretization errors are likely to be
larger than in the light hadron spectrum. The improvement of the NRQCD
Lagrangian is thus a crucial issue and its efficiency has to be
checked explicitly by simulating at different values of $\beta$. We
have removed $O(a^2)$ errors in the tree approximation, but we cannot
perform a scaling analysis for the dynamical data, as we are
restricted to a single lattice spacing.  In the quenched approximation
Davies et. al. \cite{davies98} have studied the bottomonium spectrum
on three quenched lattices with spacings in the range 0.05 fm to 0.15
fm. They find good scaling in the ratio of radial and orbital $b\bar
b$ splittings to the $\rho$-mass if the latter is computed from a
tadpole improved clover action. Also ratios of such splittings within
the $\Upsilon$-system do not exhibit any dependence on the lattice
spacings (spin splittings do!).  Although these results are quite
encouraging, we emphasize that dynamical Wilson fermions introduce
additional linear scaling violations whose size can only be safely 
estimated through simulations at different lattice spacings.

\section{PLAQUETTE COUPLING}
\label{sec:alphaP}
A value of the strong coupling now is readily obtained: compute the
expectation value of a short distance quantity on the lattice and match it
with the perturbative expansion. Obviously, this is a reasonable procedure
only, if nonperturbative effects are negligible which is definitely the case
for the simplest lattice quantity, the 1x1 Wilson loop. While the common
choice of the expansion parameter in the continuum is the $\overline{\rm
  MS}$-coupling, on the lattice it is more suitable to choose a subtraction
scheme that refers to a nonperturbative quantity like the static $Q\bar
Q$-potential \cite{susskind77}. Rather than using $\alpha_V$ itself we adopt here a
slightly modified scheme, defined in Ref. \cite{davies95} through
\begin{equation}
\label{eq:def_alpha_P}
  -\ln\langle \frac{1}{3}\Re {\rm Tr}U_{\mu\nu}\rangle
  = \frac{4\pi}{3}\alpha_P\left(\frac{3.41}{a}\right)\left[ 1 -
  \left(1.1870 + 0.0249\; N_f\right)\alpha_P\right] \; ,
\end{equation}
the rationale behind this definition of plaquette coupling, $\alpha_P$, being
a matter of convenience: one has to worry about higher order perturbative
corrections only once, when converting the lattice coupling into a standard
continuum scheme at the very end of the  analysis. Note that Eq.
\ref{eq:def_alpha_P} is valid in the chiral limit and for Wilson fermions. One
prefers to expand the logarithm of the plaquette since it converges more
rapidly than the plaquette expectation value itself. The scale $3.41/a$ is the
`average gluon momentum' in the first-order contribution to $-\ln W_{11}$
computed with the technique suggested in \cite{lepage93}.  In Table
\ref{tab:alphaP} we summarise the couplings $\alpha_P$ obtained from
Eq.(\ref{eq:def_alpha_P}) as well as the scales determined from the 1P-1S
($\bar\chi - S$) and 2S-1S ($\Upsilon' - \Upsilon$) splittings.

In the unquenched case we quote values for both $am_q = am_s /3$ and $am_q =
am_l$ to estimate the systematic error connected to the finite sea-quark mass.
Plaquette expectation values have been extrapolated accordingly. We do not
quote an error for them since it is negligible compared to the uncertainty in
the scale. Subsequently these couplings are evolved to a common scale $\mu =
9.0$ GeV using the universal two-loop $\beta$ function, see Table
\ref{tab:alpha_P_cs}. The error in the evolution is minute  as the
evolution range is very small. For instance, using just the one-loop evolution
results in a deviation of much less than 1\%.

\begin{figure}
\begin{center}
  \epsfxsize=12cm
  \epsfbox{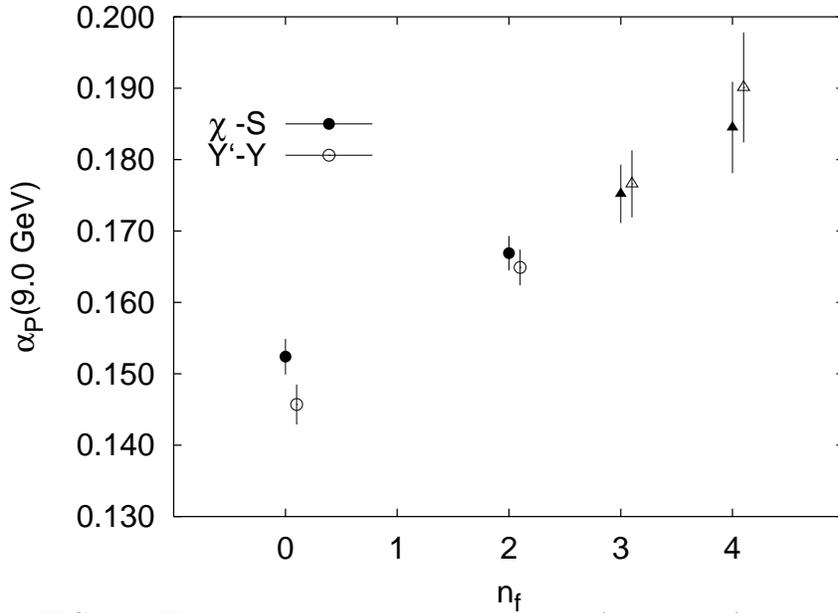}
\caption{\sl The plaquette coupling $\alpha_P$ as a function of the number
  of degenerate dynamical flavours. The triangles result from an
  extrapolation in the inverse flavour number. \label{fig:alphaP_nf}}
\end{center}
\end{figure}

The plaquette couplings in the quenched and unquenched theories can now be
extrapolated to the number of active light quark flavours which is expected to
be $N_f = 3$ in the case of the low-lying $b\bar b$ bound states. Guided by
the perturbative evolution, we extrapolate $\alpha_P^{-1}$ linearly in $N_f$,
Figure \ref{fig:alphaP_nf}. Obviously the mismatch between $\alpha_P$-values
obtained from different splittings in the quenched approximation disappears,
once the dynamical quarks are switched on. Thus we find, that full QCD with
Wilson fermions shows similar  behaviour as QCD with staggered dynamical
fermions\cite{davies95}. Nevertheless, there is no way to pin down the
the number of active flavours precisely. Continued extrapolation to
$N_f=4$ further increases $\alpha_P$ (with the error blown up) and
leads to $1\sigma$ increase in $\alpha_{\overline{\rm
    MS}}(M_Z)$. Hence simulations with $N_f > 2$ are called for to
reduce  this uncertainty.  

Consider once more the ordering of extrapolations that have been
performed to arrive at $\alpha_P^{(3)}$. First we have extrapolated
splittings in lattice units as a function of the dynamical quark mass
to determine $a^{-1}$ at $\tilde m$. Then we performed the
$N_f$-extrapolation.  One might argue about this prescription of
carrying out the limits.  Alternatively one might prefer to do the
flavour extrapolation at fixed quark mass. Although in principle at
finite lattice spacing the order may matter, we see no difference: the
couplings separately obtained at each sea-quark mass, see Table
\ref{tab:alphaP_kappa}, do not reveal any significant dependence on
$m_q$ and are consistent with those in Table \ref{tab:alpha_P_cs}.

\section{DISCUSSION}
\label{sec:discussion}
To make the connection with the $\overline{\rm MS}$-scheme one invokes
\begin{equation}
  \label{eq:conversion}
  \alpha_{\overline{\rm MS}}^{(N_f)}\left(Q\right)
  = \alpha_P^{(N_f)}\left( {\rm e}^{5/6}Q\right) \left[ 1 +
  \frac{2}{\pi}\alpha_P^{(N_f)} + C_2(N_f)(\alpha_P^{(N_f)})^2 +
  O((\alpha_P^{(N_f)})^3)\right] 
  \; ,
\end{equation}
with a scale factor ${\rm e}^{-5/6}$ chosen to eliminate the $N_f$
dependence in the first-order coefficient of the expansion \cite{brodsky83}.
The crucial point about Eq.(\ref{eq:conversion}) concerns the
coefficient $C_2$ which is only known in pure gauge theory and
thus causes a significant uncertainty on $\amsbar$. Let us, for the moment,
ignore this uncertainty and set $C_2\approx 0.95$, as in the
quenched theory, to study the errors directly related to the lattice method.
We start from $\alpha_P^{(3)}(9.0\,
{\rm GeV})$ and evolve the coupling to the Z meson mass scale, 
applying the formulae in \cite{rodrigo93}: First, $\amsbar^{(3)}({\rm
  e}^{-5/6}\cdot 9.0 {\rm GeV})$, is evolved down to the charm
threshold with the three-loop beta function. Matching the three
flavour with the four flavour theory, one obtains $\amsbar^{(4)}(M_c)$
which in turn is evolved upwards to the b-quark threshold. We have
chosen $m_c=1.3$ GeV and $m_b=4.1$ GeV for the charm and 
bottom thresholds, see Table \ref{tab:amsbar}. As was already
noted in \cite{davies96}, the value of the coupling at $m_Z$ is insensitive to
the precise location of the matching point.
Errors are propagated
by performing the evolution on each bootstrap sample separately. They
turn out to exceed the effect of this matching procedure by an order
of magnitude. 
As a result we obtain the consistent estimates
\begin{equation}
\label{eq:al_final}
\alpha_{\overline{\rm MS}}^{(5)}\left(m_Z\right) = \left\{\begin{array}{lllll} 
    0.1118 & (10)(12)(5)  &  \mbox{from} & \bar\chi - \Upsilon & \mbox{splitting}\\
    0.1124 & (13)(12)(15) &  \mbox{from} & \Upsilon ' - \Upsilon & \mbox{splitting}
  \end{array} \right.  
\end{equation} 
We give three errors to
quantify the uncertainty in the lattice scale determination: the
statistical error, the systematic error of the NRQCD expansion and the
uncertainty originating from the sea-quark mass
dependence. eq.\ref{eq:al_final} permits the following conclusions: we
have attained statistical errors on the level of the systematic
uncertainties, hence they are not the limiting factor, even in
the unquenched theory. In addition, the errors induced by applying
nonrelativistic QCD and unphysically heavy sea quarks appear to be
fairly well under control.

One might be tempted to grade eq. \ref{eq:al_final} as a ``high
precision'' determination of $\alpha_s$:
Adding the errors quadratically, one finds in fact an
overall uncertainty of ``only'' $\sim 1.5\%$, which is quite small compared
to the errors found in recent experimental measurements
of the strong coupling \cite{particle_book}.

This conclusion, however, might be misleading. Recall that our
analysis has been performed within a fixed (Wilson) discretization
scheme, at a given value of the lattice cutoff $a^{-1}$, and with 
an incomplete conversion prescription 
$\alpha_{\overline{\rm MS}}^{(N_f)}(\alpha_P^{(N_f)})$, 
c.f. eq.\ref{eq:conversion}. Clearly, systematic effects which might
arise from these limitations can be taken into account properly
only by variation of the setup.   

To estimate the size of these additional uncertainties we compare our
results with those of ref. \cite{davies95}. The latter analysis has
been done at a similar value of the lattice cutoff, with the same
conversion prescription
$\alpha_{\overline{\rm MS}}^{(N_f)}(\alpha_P^{(N_f)})$, but within
the Kogut Susskind discretization scheme. As their final result, the
authors of ref.  \cite{davies95} quote
$\alpha^{(5)}_{\overline{\rm MS}}(m_Z)$ to be 
0.1174(15) and 0.1173(21) from the $\bar\chi-\Upsilon$ and $\Upsilon '
-\Upsilon$ splitting, respectively. These values exceed our result by
5\% or three sigmas! Thus, one concludes that the ``true'' systematic
uncertainty is  3 to 4 times larger than the one given in
eq. \ref{eq:al_final}\footnote{Of course, the same error ($5\%$)
should be added to the result of ref. \cite{davies95} since, from
the current stage of knowledge, one cannot tell which 
discretization scheme yields results closer to the continuum.}.  

Let us discuss the possible origins of the additional uncertainty
in some more detail:

(1) Both, the errors caused by the unknown
flavour dependence of $C_2$ and the truncation of the perturbative
series, Eq. \ref{eq:conversion}, have been ignored up to now. To
estimate their magnitude we vary the $N_f$-dependent part of $C_2$
within a reasonable range, allowing values between -1 and +1. In
addition we set the coefficient of the third-order contribution to unity. The
latter turns out to have practically no effect, whereas the variation
of $C_2$ suggests an extra 2-3\% uncertainty.
It is thus not implausible that the discrepancy in
$\amsbar^{(5)}$ is due to our ignorance on $C_2$.

(2) The difference in couplings between Wilson and staggered
data is already present prior to converting to the continuum
renormalization scheme. We have evolved $\alpha_P^{(2)}$ as it
results from our analysis to the momentum scale used in
ref. \cite{davies95} and find $\alpha_P^{(2)}(8.2\, {\rm GeV}) =
0.1714(25)$, a 4\% difference compared to the staggered result.   
This deviation can be traced back directly to the difference in
plaquette values which is larger than anticipated from the
perturbative expansion, while the scales in
both simulations are compatible within errors. This signals the
presence of sizeable finite-a errors.

Both points constitute substantial limitations.

To reduce
the error stemming from source (1) it is of utmost importance to 
calculate perturbatively the coefficient $C_2(N_f)$ both, for Wilson
and for Kogut Susskind fermions.

A reduction of the uncertainty related to source (2) will require a
much more detailed 
 numerical analysis. Since the continuum limit 
$a\rightarrow 0$ does not exist in NRQCD, one cannot remove cutoff
effects by extrapolation in $a$. Instead one has to rely on improved
discretization schemes, which avoid sizeable cutoff effects already at
finite $a$. The compelling test however, whether a given scheme
really reduces cutoff 
effects compared to Wilson or Kogut Susskind discretizations can be
performed only by a scaling analysis in full relativistic lattice QCD. Thus,
in a sense, improvement of NRQCD presupposes the improvement of 
relativistic lattice QCD with respect to discretization errors.

\section{SUMMARY AND CONCLUSION}  Heavy quarkonium
bound states are potentially able to reveal 
the QCD coupling to high precision.  Our main objective here has been
to acquire a better understanding of the various sources of error when
applying NRQCD techniques to carry out this program in lattice QCD.
We have been able to reduce statistical errors to the level of
systematic effects. Among
the latter, uncertainties from the truncation of the NRQCD action and
the dynamical quark mass dependence are found to be under good
control. On the other hand, errors due to flavour extrapolation are
more subtle to pin down but seemingly not dominant.   
 Much more relevant is the choice of lattice action for the
light quarks. Our analysis, using Wilson quarks, leads to a value of
$\amsbar^{(5)}(m_Z)$ significantly smaller than comparable
calculations based on 
staggered light quarks. This suggests that discretization errors
do play an important role in limiting the precision
of NRQCD type determinations of the strong coupling and suggests to
 base the analysis on the use of improved Wilson type actions.
Needless to say, as a first step, one must improve the perturbative
recoupling between the lattice and $\overline{MS}$ schemes. 
\section*{ACKNOWLEDGEMENTS}

We thank the members of the Wuppertal DELPHI group for fruitful
discussions. A.S. wishes to thank C. Davies for many helpful
suggestions. We appreciate support by EU network Grant No. ERB CHRX CT
92-0051. Computations were done on the Connection Machines CM5 in Wuppertal and
Erlangen. We are grateful to the staff of the Rechenzentrum at the
University of Erlangen.  

% --- BIBLIOGRAPHY

% --- TABLES

\begin{table}
\begin{center}
\caption{\sl Simulation set-up. Where two numbers are quoted they
  refer to S-wave/P-wave correlators. \label{tab:simsetup}} 
\begin{tabular}{lllll}
& \multicolumn{3}{l}{Dynamical Wilson} & Quenched  \\
& \multicolumn{3}{l}{$\beta=5.6$ , $16^3\times 32$} &  $\beta=6.0$ ,
$16^3\times 32$ \\
\hline
$\kappa$           & 0.1560   & 0.1570    & 0.1575    & \\
$m_{\pi}/m_{\rho}$ & 0.833(3) & 0.758(11) & 0.686(11) & \\
\# configurations     & 206      & 192       & 203       & 811/520 \\
\# point sources      & 4/4      & 13/8      & 12/12     & 4/4 \\
\end{tabular}
\end{center}
\end{table}

\begin{table}
\begin{center}
\caption{\sl Fit results in lattice units. $1^3S_1$, $2^3S_1$ and
  $1^1P_1$ are obtained from two-exponential
  fits to smeared-local correlators, spin splittings are taken
  from ratio fits to smeared-smeared correlators.\label{tab:final_ro}}
\begin{tabular}{lllll}
& \multicolumn{3}{c}{$N_f = 2, \beta = 5.6$}  & $N_f = 0, \beta =
6.0$ \\
\hline
Level & $\kappa = 0.1575$ & $\kappa = 0.1570$ & $\kappa = 0.1560$ &
\\
\hline
$1\, ^3S_1$ & 0.3584(6) & 0.3606(6) & 0.3652(9) & 0.3438(4) \\
$2\, ^3S_1$ & 0.590(8)  & 0.601(10) & 0.631(20) & 0.589(12) \\
$1\, ^1P_1$ & 0.530(5)  & 0.536(6)  & 0.549(7)  & 0.508(6)  \\
\hline
$ 1^3P_2 - 1\bar P $ & 0.0034(4)  & 0.0032(4)   & 0.0028(4)  & 0.0032(3) \\
$ 1\bar P - 1^3P_1 $ & 0.0024(5)  & 0.0021(5)   & 0.0018(3)  & 0.0020(4) \\
$ 1\bar P - 1^3P_0 $ & 0.0099(10) & 0.0096(11)  & 0.0087(10) & 0.0098(5) \\
$ 1\bar P - 1^1P_1 $ & 0.0003(4)  & 0.0000(5)   & -0.0001(3) & 0.0004(3) \\
\end{tabular}
\end{center}
\end{table}  

\begin{table}
\begin{center}
\caption{\sl Results of the extrapolation in the sea-quark
  mass. $^3\bar P$ is the spin-averaged triplet-P
  state. \label{tab:extrap_results}}  
\begin{tabular}{cllll}
splitting & $a\Delta E_0$ & $c$ & \hspace{-0.1cm}$a\Delta E(m_s/3)$ &
$a\Delta E(m_l)$\\ 
\hline
$2^1S_0 - 1^1S_0$ & 0.209(21) & 1.2(7)   & 0.229(10) & 0.212(19) \\
$2^3S_1 - 1^3S_1$ & 0.209(18) & 1.1(7)   & 0.226(9)  & 0.211(17) \\
$1^3\bar P - 1^3S_1$ & 0.163(9)  & 0.4(3)   & 0.170(5) & 0.164(8)  \\
$2^1P_1 - 1^1P_1$ & 0.152(24) & 1.8(7)   & 0.181(15) & 0.156(23) \\
\end{tabular}
\end{center}
\end{table} 

\begin{table}
\begin{center}
\caption[Lattice spacings]{\sl Determination of the lattice spacing
  from the $2^3S_1-1^3S_1$ and $1^3\bar P-1^3S_1$
  splittings.\label{tab:spacings}}  
\begin{tabular}{llllll}
& $N_f = 0$ & $\kappa = 0.1560 $ & $\kappa = 0.1570 $ &
$\kappa = 0.1575 $ & $ m_s/3 $ \\
\hline
$a^{-1}(\Upsilon ' - \Upsilon)[{\sf GeV}]$ & 2.29(11) & 2.12(16) &
2.34(9) & 2.43(8) & 2.49(10) \\
$a^{-1}(\bar\chi - \Upsilon)[{\sf GeV}]$ & 2.68(9) & 2.38(8) & 2.50(8)
& 2.57(7) & 2.59(7) \\
\end{tabular}
\end{center}
\end{table}

\begin{table}
\begin{center}
\caption[Comparison of LO and NLO NRQCD results]{\sl Comparison of
  quenched $\Upsilon$ 
  and $\eta_b$ energies obtained with 
  the $O(m_b v^4)$ and $O(m_b v^6)$ actions. Simulation parameters are
  $\beta = 6.0$, 
  $M_b = 1.7$ . \label{tab:lonlo}}
\begin{tabular}{cclllll}
action & $u_0$ & $1\, ^1S_0$ & $2\, ^1S_0$ & $1\, ^3S_1$ & $2\, ^3S_1$
& $1\, ^3S_1 - 1\, ^1S_0$ \\
\hline
$O(m_b v^4)$ & $u_0^L$ & 0.3299(6) & 0.581(12) & 0.3441(7) & 0.587(14) &
0.01443(23) \\ 
$O(m_b v^6)$ & $u_0^L$ & 0.3309(4) & 0.582(12) & 0.3438(4) & 0.589(12) &
0.01266(8) \\ 
\end{tabular}
\end{center}
\end{table}

\begin{table}
\begin{center}
\caption[Results for $\alpha_P\left(3.41/a\right)$]{\sl Results for
  $\alpha_P\left(3.41/a\right)$ extracted from the 
  measured plaquette value. We quote the statistical error on the
  lattice scale as well as the systematic NRQCD error.\label{tab:alphaP}} 
\begin{tabular}{cccccll}
$\beta$ & $N_f$ & $M_q$ & $-\ln\langle \frac{1}{3}\Re{\rm Tr}
U_{\mu\nu}\rangle$ & 
$\alpha_P^{(N_f)}(\frac{3.41}{a})$ &
\multicolumn{2}{c}{$\frac{3.41}{a}$ [GeV]} \\
&&&&& $\bar\chi - \Upsilon$ & $\Upsilon^{\prime} - \Upsilon$ \\
\hline
5.6 & 2 & $M_s/3$  & 0.5570 & 0.1678 & 8.84(26)(35) & 8.48(33)(32) \\
    & 2 & $M_l$    & 0.5546 & 0.1668 & 9.16(45)(70) & 9.09(72)(71) \\
6.0 & 0 & $\infty$ & 0.5214 & 0.1518 & 9.13(34)(37) & 7.82(38)(38) \\
\end{tabular}
\end{center}
\end{table}

\begin{table}
\begin{center}
\caption[Plaquette couplings at the reference scale $\mu =
9.0$ GeV]{\sl Plaquette couplings at the reference scale $\mu = 9.0$ 
  GeV. The last two columns are the result of an extrapolation in the
  flavour number. The first error is statistical, the second is the
  systematic error within NRQCD. \label{tab:alpha_P_cs}}
\begin{tabular}{cccccc}
Splitting & $\alpha_P^{(0)}\left( 9.0 \,{\rm GeV}\right)$ &
\multicolumn{2}{c}{$\alpha_P^{(2)}\left( 9.0 \,{\rm GeV}\right)$} &
\multicolumn{2}{c}{$\alpha_P^{(3)}\left( 9.0 \,{\rm GeV}\right)$}  \\
&& $M_q = M_s/3$ & $M_q = M_l$ & $M_q = M_s/3$ & $M_q = M_l$ \\ 
\hline
$\bar\chi - \Upsilon$ & 0.1525(17)(18) & 0.1670(14)(19) & 0.1677(24)(36) &
0.1753(25)(32) & 0.1764(41)(62) \\ 
$\Upsilon^{\prime} - \Upsilon$ & 0.1458(20)(20) & 0.1650(18)(17) & 0.1673(38)(37) &
0.1767(33)(33) & 0.1806(69)(68) \\
\end{tabular}
\end{center}
\end{table}

\begin{table}
\begin{center}
\caption[Plaquette couplings at the reference scale $\mu =
9.0$ GeV]{\sl Plaquette couplings at the reference scale $\mu = 9.0$ 
  GeV for each sea quark mass. Here, we quote only one error, covering
  both statistical and systematic uncertainties.\label{tab:alphaP_kappa}}
\begin{tabular}{ccccccc}
$\kappa$ & $\langle\frac{1}{3}\Re{\rm Tr} U_{\mu\nu}\rangle$ &\multicolumn{2}{c}{$\bar\chi - \Upsilon$} &
\multicolumn{2}{c}{$\Upsilon^{\prime} - \Upsilon$} \\
&& $\alpha_P^{(2)}\left( 9.0 \,{\rm GeV}\right)$ &
$\alpha_P^{(3)}\left( 9.0 \,{\rm GeV}\right)$ & $\alpha_P^{(2)}\left(
  9.0 \,{\rm GeV}\right)$ &   
$\alpha_P^{(3)}\left( 9.0 \,{\rm GeV}\right)$ \\ 
\hline
0.1560 & 0.5698 & 0.1651(26) & 0.1723(46) & 0.1598(39) & 0.1679(68) \\
0.1570 & 0.5716 & 0.1661(25) & 0.1739(43) & 0.1631(26) & 0.1733(55) \\
0.1575 & 0.5725 & 0.1667(22) & 0.1749(37) & 0.1642(26) & 0.1753(44) \\
\end{tabular}
\end{center}
\end{table}

\begin{table}
\begin{center}
\caption[$\overline{\rm MS}$ coupling at the heavy-quark thresholds and
  the Z pole]{\sl $\overline{\rm MS}$ coupling at the heavy-quark
  thresholds and the Z mass.\label{tab:amsbar}}
\begin{tabular}{ccccccc}
Splitting & \multicolumn{2}{c}{$\amsbar\left( 1.3 \,{\rm GeV}\right)$} &
\multicolumn{2}{c}{$\amsbar\left( 4.1 \,{\rm GeV}\right)$} &
\multicolumn{2}{c}{$\amsbar\left( 91.2 \,{\rm GeV}\right)$}  \\
$M_q$  & $M_s/3$ & $M_l$ & $M_s/3$ & $M_l$ & $M_s/3$ & $M_l$\\  
\hline
$\bar\chi - \Upsilon$       & 0.316(9)(11) & 0.320(14)(22) & 0.2034(33)(44) &
0.2050(55)(84) & 0.1118(10)(12) & 0.1123(16)(24) \\  
$\Upsilon^{\prime} - \Upsilon$ & 0.321(12)(9) & 0.335(26)(26) & 0.2053(45)(37)  &
0.2106(95)(89) & 0.1124(13)(12) & 0.1139(27)(28) \\
\end{tabular}
\end{center}
\end{table}

% --- FIGURES

\end{document}